\documentstyle[12pt]{article}
\newcommand{\refer}[1]{\vspace{-3mm}\bibitem{#1}}
\newcommand{\bz}{\begin{itemize}}
\newcommand{\ez}{\end{itemize}}
\renewcommand{\Im}{\II\!m\,}
\renewcommand{\Re}{\RR\!e\,} 
\newcommand{\point}[1]{\vspace{1mm} 

\noindent#1}
\newcommand{\remark}{\vspace{1mm} 

\noindent{\bf Remark:\ }}
\newcommand{\kramer}{\vspace{1mm} 

}
\newcommand{\remarks}{\vspace{2mm} 

\noindent{\bf Remarks:}}
\newcommand{\theorem}{\vspace{2mm} 

\noindent{\bf Theorem.\ }}
\def\bbbc{{\mathchoice {\setbox0=\hbox{$\displaystyle\rm C$}\hbox{\hbox
to0pt{\kern0.4\wd0\vrule height0.9\ht0\hss}\box0}}
{\setbox0=\hbox{$\textstyle\rm C$}\hbox{\hbox
to0pt{\kern0.4\wd0\vrule height0.9\ht0\hss}\box0}}
{\setbox0=\hbox{$\scriptstyle\rm C$}\hbox{\hbox
to0pt{\kern0.4\wd0\vrule height0.9\ht0\hss}\box0}}
{\setbox0=\hbox{$\scriptscriptstyle\rm C$}\hbox{\hbox
to0pt{\kern0.4\wd0\vrule height0.9\ht0\hss}\box0}}}}

\newcommand{\bi}{\begin{itemize}} \newcommand{\ei}{\end{itemize}}

\newcommand{\bint}[2]{{\dint_{\kern -.4em #1}^{#2}}}

\newcommand{\Bsurl}[1]{ 
{\raise 5ex \hbox{$\overline{{\raise -5ex \hbox{$#1$} }}$} }
}
\newcommand{\bsurl}[1]{ 
{\raise 3ex \hbox{$\overline{{\raise -3ex \hbox{$#1$} }}$} }
}
\newcommand{\surl}[1]{ 
{\raise 2ex \hbox{$\overline{{\raise -2ex \hbox{$#1$} }}$} }
}

\newcommand{\ad}{{\dot\a}}

\newcommand{\cb}{{\bar{c}}}
\newcommand{\CB}{{\bar{C}}}

\newcommand{\pbar}{{\bar\p}}

\newcommand{\lb}{{\bar\la}}

\def\ftoday{{\sl  \number\day \space\ifcase\month 
\or Janvier\or F\'evrier\or Mars\or avril\or Mai
\or Juin\or Juillet\or Ao\^ut\or Septembre\or Octobre
\or Novembre \or D\'ecembre\fi
\space  \number\year}}    
\newcommand{\journal}[4]{{\em #1~}#2\,(19#3)\,#4;}

\newcommand{\hpa}{\journal {Helv. Phys. Acta}}

\newcommand{\ijmp}{\journal {Int. J. Mod. Phys.}}

\newcommand{\pr}{\journal {Phys. Rev.}}

\newcommand{\cmp}{\journal {Commun. Math. Phys.}}

\newcommand{\cqg}{\journal {Class. Quantum Grav.}}

\newcommand{\np}{\journal {Nucl. Phys.}}
\newcommand{\pl}{\journal {Phys. Lett.}}
\newcommand{\mpl}{\journal {Mod. Phys. Lett.}}

\newcommand{\es}{\\[2mm]}

\renewcommand{\a}{\alpha}
\renewcommand{\b}{\beta}
\newcommand{\g}{\gamma}           
\renewcommand{\d}{\delta}         
\newcommand{\e}{\varepsilon}

\newcommand{\la}{\lambda}        
\newcommand{\m}{\mu}
\newcommand{\n}{\nu}
         
\newcommand{\p}{\psi}              
\newcommand{\r}{\rho}
\newcommand{\s}{\sigma}           \renewcommand{\S}{\Sigma}
\newcommand{\th}{\theta}         
\newcommand{\f}{{\phi}}           

\newcommand{\II}{{\cal I}}
\newcommand{\JJ}{{\cal J}}

\newcommand{\SS}{{\cal S}}
\newcommand{\RR}{{\cal R}}

\newcommand{\xint}{\dint d^4x\;}
\newcommand{\sla}{\raise.15ex\hbox{$/$}\kern -.57em} 
\newcommand{\Sla}{\raise.15ex\hbox{$/$}\kern -.70em}
\def\h{\hbar}

\def\LP{\displaystyle{\Biggl(}}
\def\RP{\displaystyle{\Biggr)}}
\newcommand{\lp}{\left(}\newcommand{\rp}{\right)}

\newcommand{\lac}{\left\{}\newcommand{\rac}{\right\}}

\newcommand{\complex}{{\kern .1em {\raise .47ex
\hbox {$\scriptscriptstyle |$}}
    \kern -.4em {\rm C}}}
\newcommand{\real}{{{\rm I} \kern -.19em {\rm R}}}
\newcommand{\rational}{{\kern .1em {\raise .47ex
\hbox{$\scripscriptstyle |$}}
    \kern -.35em {\rm Q}}}
\renewcommand{\natural}{{\vrule height 1.6ex width
.05em depth 0ex \kern -.35em {\rm N}}}

\newcommand{\half}{\dfrac{1}{2}}
\newcommand{\pa}{\partial}

\newcommand{\dfrac}[2]{{\displaystyle{\frac{#1}{#2}}}}
\newcommand{\dsum}[2]{\displaystyle{\sum_{#1}^{#2}}}   
\newcommand{\dint}{\displaystyle{\int}}

\newcommand{\twiddle}{\lower.9ex\rlap{$\kern -.1em\scriptstyle\sim$}}

\newcommand{\equ}[1]{(\ref{#1})}
\newcommand{\eq}{\begin{equation}}
\newcommand{\eqn}[1]{\label{#1}\end{equation}}
\newcommand{\eea}{\end{eqnarray}}
\newcommand{\eqa}{\begin{eqnarray}}
\newcommand{\eqan}[1]{\label{#1}\end{eqnarray}}
\newcommand{\ba}{\begin{array}}
\newcommand{\ea}{\end{array}}
\newcommand{\eqac}{\begin{equation}\begin{array}{rcl}}
\newcommand{\eqacn}[1]{\end{array}\label{#1}\end{equation}}

\begin{document}
\noindent{hep-th/9606045\\ UGVA-DPT-1996/06-931
 \vspace{8mm}
\begin{center}
{\Large{\bf Supersymmetry, Ultraviolet Finiteness\\[3mm]
 and Grand Unification}}\footnote{Talk given at the International conference
``Problems of Quantum Field Theory'', 
Alushta (Crimea, Ukraine, May 1996).}
\vspace{6mm}

{\large Olivier Piguet}\footnote{Supported in part 
                 by the Swiss National Science Foundation.}
\vspace{3mm}

\noindent {\small D\'epartement de Physique Th\'eorique, 
                                  Universit\'e de Gen\`eve\\
          24, quai Ernest Ansermet,
          CH -- 1211 Gen\`eve 4 (Switzerland)}
\vspace{10mm}

\end{center}
\vspace{2mm}

\noindent {\small {\em Abstract.\ }      
A general criterion is given for the vanishing of the $\b$-functions in
$N$=1 supersymmetric gauge theories.}
\subsection*{1 \ Introduction and Conclusions}
Supersymmetry is well known to enforce cancellations of ultraviolet (UV)
divergences. $N$=4 super-Yang-Mills (SYM) theory for instance is
devoid
of any UV divergence~\cite{sohnius-west,mandelstam,white2}, 
and this holds as well for a class of $N$=2 SYM
models~\cite{howe-stelle-west-town}. 
Our aim is to present similar 
results~\cite{beta0,beta0massless,claudio} for
$N$=1 SYM theories in 4-dimensional spacetime. We
mean here, by UV finiteness, the vanishing of all the
$\b$-functions, i.e. the nonrenormalization of the coupling
constants. To the contrary of other
approaches~\cite{jones-kazakov,kazakov} 
which tend to complete UV finiteness, we don't require the (nonphysical) 
anomalous dimensions to vanish, i.e. infinite (unobservable) field amplitude
renormalizations may still be present. 

The physical interpretation of our results is scale invariance -- or
better, asymptotic scale invariance since masses may be present-- but we
shall consider the massless case~\cite{beta0massless}
for the sake of simplicity.

We shall give a general criterion, which involves
one-loop quantities only,
for the vanishing of the $\b$-functions at all orders of perturbation
theory
The criterion
is based, first, on a relation between the anomaly of the
axial $R$-current 
and the scale anomaly -- expressed by the $\b$-functions -- which
follows from the axial current and the energy-momentum tensor
belonging to the supercurrent 
multiplet of Ferrara and Zumino~\cite{fer-zu},
and from the nonrenormalization
theorem for the axial anomaly.
The second ingredient is the fact that
the Yukawa couplings must necessarily be functions of the gauge coupling
constant, functions solving the reduction equations of
Oehme and Zimmermann~\cite{oeh-zi}. 
The criterion is quite
general, it is independent of the renormalization scheme used and 
it does not rely on the existence of a regularization preserving
both gauge invariance and supersymmetry\footnote{In a generic scheme the
Yukawa $\b$-functions are not necessarily linear combinations of the
matter anomalous dimensions -- except in the one-loop approximation --
but this has no consequence on our result, which concerns the
$\b$-functions only.}. 
The procedure is based on
general results of renormalization theory~\cite{pig-sor}.

The only restriction is the assumption that the gauge group is a simple Lie
group. But this lets it remain very interesting in the framework of
Grand Unified Theories, where it may lead to predictions for the mass
spectrum in particular.
Applications of the criterion may be found in the second of 
refs.~\cite{beta0} for a SU(6) model and, more interestingly,
in~\cite{zoup,claudio}, for a realistic SU(5) model compatible with
the Minimal Supersymmetric Standard Model at low energies, which
predicts a top mass of $(185\pm5)$ GeV (see also~\cite{kazakov}).

Our criterion can actually be considered as
the rigorous version of a formal argument 
already given by the authors of 
Ref.~\cite{sohnius-west} for the case of $N$ = 4 SYM.

Let us finally mention that we restrict ourselves here to
unbroken symmetry. The case of supersymmetry being broken by soft mass
terms is under study~\cite{mpw} in the framework of the Wess-Zumino gauge.
A superspace approach with complete UV finiteness is proposed 
in~\cite{kazakov}. 
 
\subsection*{2 \ Super Yang-Mill Theory}
A generic $N$=1 supersymmetric gauge theory, with a simple Lie group 
$G$ as gauge group, is given at the classical level by
\point{1.}
Supermultiplets $V^a$ of gauge fields, each containing in particular 
a gauge vector field $A^a_\m$
and a gaugino Weyl spinor field  $\la^a_\a$, in the adjoint representation
of the gauge group, as well as matter
supermultiplets $S_i$, each containing in particular 
a scalar field $\f_i$ and a Weyl spinor
field $\p_{i\a}$, in some unitary representation $R$. 
The gauge fixing is implemented through 
Lagrange multiplier chiral supermultiplets $B^a$ and 
ghost (antighost) chiral supermultiplets $C_+^a$ ($\CB_-^a$) containing
in particular the ordinary ghosts (antighosts) $c^a$ ($\cb^a$).
External fields (the ``antifields'') $A^{*a}_\m$, $\la^{*a}_\a$, 
$\f^{*i}$, $\p^{*i}_\a$, etc. have to be introduced in order to
control the renormalization
of the BRS transformations given below, which are not linear.
\point{2.} The BRS transformations
\eq\ba{ll}
s A^a_\m = (D_\m c)^a +\cdots := \pa_\m c+ f^{abc}A^b_\m c^c +\cdots\ ,
 \quad& s \la^a_\a = -f^{abc}c^b\la^c_\a +\cdots\ ,\es
s \f_i = -{{R_a}_i}^j c^a \f_j +\cdots\ , 
 &s \p_{i\a} = -{{R_a}_i}^j c^a \p_{j\a} +\cdots\ , \es
s c^a = -\half f^{abc} c^b c^c +\cdots\ ,&\cdots \ .
\ea\eqn{brs-transf}
\point{3.} The BRS invariant action
\eq\ba{l}
\S=\xint\LP -\dfrac{1}{4g^2} F^{a\m\n}F_{a\m\n} 
    -\dfrac{i}{g^2} \la^{a\a}\s^\m_{\a\ad}D_\m\lb^\ad_a 
    + \overline{D^\m\f_i}D_\m\f^i 
                - i \p^{i\a}\s^\m_{\a\ad}D_\m\pbar^\ad_i \es
\phantom{\S=\xint\LP} +\la_{(ijk)}\p^i\p^j\f^k + \mbox{conj} 
  \quad + \cdots \RP \ ,
\ea\eqn{class-action}
where $g$ is the gauge coupling constant and the symmetric invariant
tensors $\la_{ijk}$ are the Yukawa coupling constants.
We take both the gauge and the matter fields massless.
\kramer
\remark  
Our notations are very sketchy, the details may be found in the original
literature~\cite{ps-book,beta0}.
First, we have written everything in terms of component fields
instead of
superfields for the sake of transparency. Second, we have omitted 
the contributions of a 
lot of component fields, namely the auxiliary fields, the
nonphysical components of the gauge superfield, as well as 
the Lagrange multiplier, ghost and antighosts fields, and the
external fields. All these omissions are signalized by dots in 
\equ{brs-transf}, \equ{class-action} and in the following.
Third, many numerical coefficients have been skipped and 
arbitrarily replaced by the number 1.
\subsection*{3 \ The Scale Anomaly}
The theory being massless is scale invariant. But this is generally true
only in the classical approximation. Radiative corrections cause a
breaking of this invariance, i.e. there is a scale anomaly. It is well
known that the scale anomaly manifests itself as a nonvanishing trace of
the energy-momentum tensor $T_{\m\n}$ -- which may be assumed to be
traceless in the tree (i.e. classical)
approximation\footnote{We expand in the powers of
Planck's constant $\h$, i.e. in the number of loops in Feynman graphs.}: 
\eq
T^\m_\m = \b_g\lp F^{\m\n}F_{\m\n}+\cdots\rp +
  \dsum{ijk}{}\lp \b_{ijk}\p^i\p^j\f^k +\cdots\rp + \cdots+ O(\h^2)\ 
,
\eqn{trace-anom}
where
\eq\ba{ll}
\b_g = \h b_0 +O(\h^2)\ ,\quad  
            &b_0 :=  l(R)-3C_2(G) \ ,\es
\b_{ijk} = \h b_{ijk} + O(\h^2)\ ,\quad
   &b_{ijk} :=    \dsum{{\rm cycl}\,(ijk)}{}
  \la_{ijn}\lp \la^{npq}\la_{pqk}-2\d^n_k g^2C_2(R) \rp 
\ea\eqn{beta-fcts}
are the $\b$-functions corresponding to the renormalizations of the
gauge coupling $g$ and of the Yukawa couplings $\la_{ijk}$, respectively. 
$l(R)$ is the Dynkin index of the representation $R$.
We have not explicitly written the contributions of the anomalous
dimensions of the various fields.
\remark
Let us recall that the $\b$-functions determine the behaviour of the
effective coupling constants as solutions of the differential equations 
\eq  
\m\dfrac{d}{d\m} \bar g(\m) = \b_g(\m,\bar g,\lb)\ ,\quad
\m\dfrac{d}{d\m} \lb_{ijk}(\m) = \b_{ijk}(\m,\bar g,\lb)\ ,
\eqn{eff-couplings}
where $\m$ is the energy scale.
\kramer
\subsection*{4 \ $R$-Invariance and Axial Anomaly}
A commun features of massless $N$=1 supersymmetric theories is
their invariance under the U(1) chiral transformation $R$:
\eq
A'_\m =   A_\m \ ,\quad\la'_\a =  e^{-i\th}\la_\a \ ,\quad
\f' =  e^{-i\frac{2}{3}\th}\f \ , \quad 
\p'_\a =  e^{i\frac{1}{3}\th}\p_\a \ ,\quad \cdots\ .
\eqn{r-transf}
The associated axial Noether current $J_R^\m(x)$
\eq
 J^\m_R = \bar\la \g^\m\g^5\la + \cdots \quad\mbox{(Dirac notation)}
\eqn{R-current}
is conserved in the classical limit, but not in the quantum 
case due to the axial anomaly:
\eq
\pa_\m J^\m_R = r \lp \e^{\m\n\r\s}F_{\m\n}F_{\r\s} +\cdots\rp
\eqn{axial-anom}
The anomaly coefficient $r$ has the remarkable property of being
equal to the one-loop value of the gauge $\b$-function 
(see \equ{beta-fcts}):
\eq
r = \h  b_0\ .
\eqn{anom-coeff}
Moreover, due to a nonrenormalization 
theorem~\cite{axial-nonren,lps}~\footnote{see~\cite{beta0} 
for the generalization to the present 
case of
supersymmetric gauge theories.}, the one-loop value given here is 
exact, without any higher order contributions.
\subsection*{5 \ The Supercurrent}
We shall now see that the equality \equ{anom-coeff} between the 
$R$-anomaly coefficient and the one-loop value of the gauge 
$\b$-function is not an 
accident. It is actually linked to the existence of a 
supermultiplet of gauge invariant operators: the 
supercurrent~\cite{fer-zu} $\JJ$:
\eq
\JJ := \lac {J'}^\m_R\ ,\ \ {Q^\mu}_\a\ ,\ \ {T^\m}_\n \rac\ ,
             \ \ \cdots\ ,
\eqn{super-current}
made of the (classically)
conserved currents associated to $R$ invariance, 
supersymmetry and translation invariance, respectively. Note that 
the latter two remain conserved to all orders, to the contrary 
of the first one. Note also that 
the $R$-current ${J'}_R^\m$ in \equ{super-current} differs from  
the $R$-Noether current \equ{R-current}. But both coincide
in the tree approximation: ${J'}_R^\m = J^\m_R + O(\h)$.
The point is that, starting from the unique classical Noether 
current $J^\m_{R\,{\rm (class)}}$, $J^\m_R$ is defined as the 
quantum 
extension which allows for the validity of a nonrenormalization
theorem~\cite{lps}, whereas ${J'}_R^\m$ is defined to belong 
together with the energy-momentum tensor to one
supermultiplet -- the supercurrent \equ{super-current}. These two 
requirements cannot be fulfilled simultaneously by a single 
current operator.

More interesting, there is a second supermultiplet (a chiral    
multiplet) containing, among others, the anomalies of the
$R$-current ${J'}_R^\m$ as well as 
the trace anomalies of the supersymmetry current and of 
the energy-momentum tensor. One can indeed prove the
set of equations -- which constitute the ``supertrace identity'':
\eq
{T^\m}_\m = \Re S\ ,\quad \pa_\m {J'}^\m_R = \Im S\ ,\quad
\s^\m_{\a\dot\b}{{\bar Q}_\m}^{\dot\b} = S_\a \ ,
\eqn{s-trace-eq}
where the anomalies in the right-hand sides belong to the chiral 
supermultiplet
\eq\ba{l}
\SS =  \lac \Re S\ ,\ \ \Im S\ ,\ \ S_\a\ ,\ \ \cdots \rac 
:= 
\es
\phantom{\SS}
\lac     \b_g F^{\m\n}F_{\m\n} +\cdots\ ,
     \ \ \b_g \e^{\m\n\r\s}F_{\m\n}F_{\r\s} +\cdots\ ,
     \ \ \b_g \la^\b\s^{\m\n}_{\a\b}F_{\m\n} +\cdots\ ,
     \ \ \cdots  \rac
\ea\eqn{s-trace-anom}
called the supertrace anomaly.
\subsection*{6 \ The Relation Between the Scale and the Axial 
 Anomalies}
Despite of the discrepancy between the axial current $J_R^\m$ obeying the
anomalous conservation law \equ{axial-anom} with a nonrenormalized
coefficient $r$ and the axial current ${J'}_R^\m$ 
belonging to the supercurrent
multiplet \equ{super-current}, whose anomaly coefficient is the
$\b$-function characterizing the scale anomaly, there is a
relation between them~\cite{beta0}:
\eq
r = \b_g (1+x_g) + \b_{ijk} x^{ijk} - \g_A r^A\ .
\eqn{relation-r-beta}
This equation relates $r$ and $\b_g$ together with the Yukawa
$\b$-functions and some other coefficients:
\point{} the $r^A$'s, which are 
the {\em nonrenormalized}  coefficients of the 
anomalies of the
Noether currents associated to the chiral invariances of the superpotential,
\point{}  the $\g_A$'s, which  are some linear combinations of 
the anomalous dimensions of the matter fields,
\point{} $x_g$ and the $x^{ijk}$'s, which are radiative 
correction 
quantities we don't need to specify.
\kramer
Let us emphasize that all the coefficients 
appearing in \equ{relation-r-beta}
are of order $\h$ at least. Moreover, $r$ and the $r^A$'s are strictly
proportional to $\h$, i.e. strictly one-loop quantities due to the
nonrenormalization theorems for the axial anomalies.
An important remark is that the structure of this identity is 
{\it independent from the renormalization scheme}, although the individual
coefficients -- except the one-loop values of the $\b$-functions -- may
be scheme dependent.

\subsection*{7 \ Ultraviolet Finiteness}
It is clear from the second of eqs. \equ{beta-fcts} that the vanishing
of the $\b$-functions implies already at the the one-loop approximation
that the Yukawa coupling constants $\la_{ijk}$ 
must be functions of the gauge coupling constant $g$. A necessary and
sufficient condition for the existence of a similar relation to all
orders is that the Yukawa coupling constants be formal power series
$\la_{ijk}(g)$, in $g$, of the reduction equations\cite{oeh-zi}
\eq
\b_g \dfrac{d\la_{ijk}}{dg} = \b_{ijk}\ .
\eqn{red-eq}
The identity \equ{relation-r-beta} then 
allows us to give a general criterion
for the ultraviolet finiteness in the sense of vanishing $\b$-functions,
i.e. of physical scale invariance:
\theorem
Consider an $N$=1 super-Yang-Mills theory with simple gauge group. If
\point{(i)} there is no gauge anomaly,
\point{(ii)} the gauge $\b$-function vanishes at one loop 
(see \equ{beta-fcts}):
\eq
b_0 := l(R)-3C_2(G) = 0\ ,
\eqn{b-g-zero}
\point{(iii)} there exist solutions of the form 
\eq
\la_{ijk}=\r_{ijk}g\ ,\quad \r_{ijk}\in\bbbc\ ,
\eqn{1-loop-red}
to the equations
\eq
\la^{jpq}\la_{pqk}-2\d^j_k g^2C_2(R) = 0\ ,
\eqn{gammazero}
\point{(iv)} These solutions are isolated and non-degenerate when
considered as solutions of vanishing one-loop Yukawa $\b$-functions 
(see \equ{beta-fcts}):
\eq
b_{ijk}=0\ ,
\eqn{b-Y-zero}
\kramer
\noindent{\it then} each of the solutions  
\equ{1-loop-red} can be uniquely
extended to a formal power series in $g$, and the associated super-YM
models depend on the single coupling constant $g$ with a $\b$-function
which vanishes at all orders.
\remarks
\point{(a)} Hyp. (ii) is equivalent to the vanishing of 
the $R$-current anomaly (see \equ{anom-coeff}). 
\point{(b)} The expressions in \equ{gammazero} 
(Hyp. (iii)) are the anomalous
dimensions of the matter fields in the one-loop approximation. Their
vanishing implies the vanishing of the one-loop Yukawa $\b$ functions
due to the second of eqs. \equ{beta-fcts}. In fact Hyp. (iii) also
implies the vanishing of the chiral anomaly coefficients $r^A$ appearing
in \equ{relation-r-beta}. The latter property is moreover a necessary
condition for having $\b$-functions vanishing to all orders.
\point{(c)} Hyp. (iv) is a condition which guaranties the existence of
a formal power series solution to the reduction equations \equ{red-eq}.
\point{(d)} It is shown in~\cite{parkes-west} that the hypotheses 
(i) to (iii) assure the vanishing of the $\b$-functions in the 
two-loop approximation. Thanks to Hyp. (iv) we are able to extand 
the result to all orders.
\kramer
\vspace{1mm} 

\noindent{\bf Proof of the theorem:\ }
Inserting $\b_{ijk}$ as given by the r.h.s. of the 
reduction equations \equ{red-eq}
into the identity \equ{relation-r-beta} and taking into account the
vanishing of the chiral anomalies $r$ and $r^A$, we get for $\b_g$ 
an homogenous equation of the form
\eq
0=\b_g\lp 1+O(\h)\rp\ .
\eqn{hom-eq}
Its solution in the sense of the formal power series in $\h$ is
$\b_g=0$. Hence $\b_{ijk}=0$ as well, due to \equ{red-eq}.
{\hfill$\Box$}
\newpage\subsection*{Discussion}
{\it Slavnov}:  Does a solution to the reduction equations always
correspond to a symmetry giving the same relation between the 
coupling constants? This seems to me necessary for the
stability of the solution.
\point{}{\it O.P.}: This may happen in the case where
the field content is compatible with a higher symmetry, like $N$=4
supersymmetry~\cite{beta0}. But I hardly see such a symmetry at work in
the SU(5) and SU(6) models mentioned in the Introduction. 
\point{}{\it Kazakov}: These solutions are infrared stable but
ultraviolet unstable.
\point{}{\it Kazakov}: Do you rely on some invariant regularization?
\point{}{\it O.P.}: No (see the Introduction).  
\medskip

{\bf Acknowledgements}: 
The author would like to manifest his gratitude 
to the organizers of
this conference for the invitation, for their very kind hospitality
and for their help for solving various travelling problems. 

\end{document}